\documentclass[review,number,sort&compress]{elsarticle}
\usepackage{tikz}
\usepackage{float}
\usepackage{graphicx}
\usepackage{gensymb}
\usepackage{amssymb}
\usepackage{color,soul}
\usepackage{url}
\usepackage{ulem}
\usepackage{textcomp}
\usepackage[figuresright]{rotating}
\usepackage{subcaption}
\usepackage{lineno}
\usepackage{hyperref}
\usepackage{cleveref}
\usepackage{libertine}
\usepackage{siunitx}
\usepackage[a4paper, total={5.0in, 8in}]{geometry}
\usepackage[T1]{fontenc}

\definecolor{amethystbg}{rgb}{0.6, 0.4, 0.8}
\definecolor{coolgreybg}{rgb}{0.55, 0.57, 0.67}
\definecolor{babypinkbg}{rgb}{0.96, 0.76, 0.76}
\definecolor{cadmiumgreenbg}{rgb}{0.0, 0.42, 0.24}
\definecolor{bluebg}{rgb}{.63,.79,.95}
\definecolor{orangebg}{rgb}{1,0.5,0}
\colorlet{lightbluebg}{bluebg!40}
\colorlet{lightorangebg}{orangebg!40}
\colorlet{lightcadmiumgreenbg}{cadmiumgreenbg!40}
\colorlet{lightbabypinkbg}{babypinkbg!40}
\colorlet{lightcoolgreybg}{coolgreybg!40}
\colorlet{lightamethystbg}{amethystbg!40}

\DeclareGraphicsExtensions{.eps,.pdf,.png,.jpg,}

\journal{Nuclear Instruments and Methods A}

\begin{document}

\newcommand{\etal}{{\it et~al.}}
\newcommand{\geant} {{{G}\texttt{\scriptsize{EANT}}4}}
\newcommand{\srim} {\texttt{SRIM}}
\newcommand{\python} {\texttt{Python}}
\newcommand{\pandas} {\texttt{pandas}}
\newcommand{\SciPy} {\texttt{SciPy}}
\newcommand{\ROOT} {\texttt{ROOT}}

\DeclareRobustCommand{\hlb}[1]{{\sethlcolor{lightbluebg}\hl{#1}}}
\DeclareRobustCommand{\hlo}[1]{{\sethlcolor{lightorangebg}\hl{#1}}}
\DeclareRobustCommand{\hlg}[1]{{\sethlcolor{lightcadmiumgreenbg}\hl{#1}}}
\DeclareRobustCommand{\hlp}[1]{{\sethlcolor{lightbabypinkbg}\hl{#1}}}
\DeclareRobustCommand{\hlgr}[1]{{\sethlcolor{lightcoolgreybg}\hl{#1}}}
\DeclareRobustCommand{\hla}[1]{{\sethlcolor{lightamethystbg}\hl{#1}}}

\begin{frontmatter}


\title{Technique for the measurement of intrinsic pulse-shape discrimination for organic scintillators using tagged neutrons}

	\author[lund]{N.~Mauritzson}
	\author[lund,ess]{K.G.~Fissum\corref{cor1}}
	\author[glasgow]{J.R.M.~Annand}
	\author[lund]{H.~Perrey}
	\author[lund]{R.J.W.~Frost}
    \author[ess,glasgow]{R.~Al~Jebali}
    \author[ess,glasgow]{A.~Backis}
	\author[ess,glasgow,milan]{R.~Hall-Wilton}
	\author[ess]{K.~Kanaki}
	\author[lund,ess]{V.~Maulerova-Subert}
	\author[lund]{C. Maurer}
	\author[lund]{F.~Messi\fnref{fn3}}
	\author[lund]{E.~Rofors}
	
	\address[lund]{Division of Nuclear Physics, Lund University, SE-221 00 Lund, Sweden}
	\address[ess]{Detector Group, European Spallation Source ERIC, SE-221 00 Lund, Sweden}
	\address[glasgow]{School of Physics and Astronomy, University of Glasgow, Glasgow G12 8QQ, Scotland, UK} 
	\address[milan]{Dipartimento di Fisica ``G. Occhialini'', Universit\`a degli Studi di Milano-Bicocca, Piazza della Scienza 3, 20126 Milano, Italy}

	\cortext[cor1]{Corresponding author. Telephone:  +46 46 222 9677; Fax:  +46 46 222 4709}        \fntext[fn2]{present address: CERN, European Organization for Nuclear Research, 1211 Geneva, Switzerland and Hamburg University, D-20148 Hamburg, Germany}
    \fntext[fn3]{present address: DVel AB, Scheelevägen 32, SE-223 63 Lund, Sweden}

\begin{abstract}
Fast-neutron/gamma-ray pulse-shape discrimination has been performed for the organic liquid scintillators NE~213A and EJ~305 using a time-of-flight based neutron-tagging technique and waveform digitization on an event-by-event basis. Gamma-ray sources and a Geant4-based simulation were used to calibrate the scintillation-light yield. The difference in pulse shape for the neutron and gamma-ray events was analysed by integrating selected portions of the digitized waveform to produce a figure-of-merit for neutron/gamma-ray separation. This figure-of-merit has been mapped as a function of detector threshold and also of neutron energy determined from time-of-flight. It shows clearly that the well-established pulse-shape discrimination capabilities of NE~213A are superior to those of EJ~305. The extra information provided by the neutron-tagging technique has resulted in a far more detailed assessment of the pulse-shape discrimination capabilities of these organic scintillators.
\end{abstract}

\end{frontmatter}
 
\newpage
\section{Introduction}
\label{section:Introduction}

Organic liquid scintillators are commonly used to detect fast neutrons in fields of gamma-rays. The main scintillation decay time constant for organic materials is generally of the order of a few ns. Several organics also have much longer decay components. The fast components are preferentially excited by relativistic particles, which are close to minimum ionizing, while slower components are excited by non-relativistic particles which ionize more heavily along their tracks. For example, NE~213A has 3 components with mean decay times of 3.2, 32.3, and 270\,ns. Secondary electrons produced by gamma-rays are close to minimum ionizing and thus give a fast signal. Protons produced after neutron scattering from scintillator hydrogen are heavily ionizing and produce more of the slow scintillation components. Thus, analysis of the fall time of the scintillation signal may be used to differentiate incident gamma-rays from neutrons. This technique is known as pulse-shape discrimination (PSD).

Due to excellent PSD capabilities, the organic liquid scintillator NE~213~\cite{ne213} and the more recent derivative NE~213A~\cite{ANNAND1997} have long been used widely~\cite{BATCHELOR196170}, and provide a performance benchmark for newly developed fast-neutron detector materials~\cite{BAYAT2012217,IWANOWSKA201334,PAWELCZAK201321,JEBALI2015102}. A more recent organic liquid EJ~305~\cite{ej305} developed for a very high scintillation-light yield is anticipated to have poorer PSD capabilities. In this paper, tagged neutrons from $\sim$1.5~--~6\,MeV are employed to investigate the PSD capabilities of these neutron-sensitive scintillators. Evaluation of the PSD performance is performed both with and without the neutron-tagging technique, and in a situation where PSD is not optimum.

\section{Apparatus}
\label{section:Apparatus}

\subsection{PuBe-based neutron and gamma-ray mixed-field source}
\label{subsection:ApparatusPuBeSource}

Fast neutrons were provided by a $^{238}$Pu/$^{9}$Be (PuBe) source. $^{238}$Pu decays to $^{234}$U via $\alpha$-particle emission (14 branches, weighted mean energy 5.4891\,MeV~\cite{nudat}), and almost simultaneously, a cascade of low-energy gamma-rays results from the de-excitation of $^{234}$U. Fast neutrons produced by the reaction $\alpha$~$+$~$^{9}$Be~$\rightarrow$~$^{12}$C~$+$~n have a maximum kinetic energy of $\sim$11\,MeV. When the final-state $^{12}$C is left in the first-excited state (with $\sim$50\% probability), it promptly de-excites by the isotropic emission of a 4.44\,MeV gamma-ray. The total radiation field associated with the PuBe source thus consists of a low-energy cascade gamma-rays, fast neutrons with energies up to $\sim$11 MeV, and a sub-field of particular interest: 4.44\,MeV gamma-rays in coincidence with fast neutrons of energy up to $\sim$6\,MeV. This sub-field of was used to produce tagged-neutron beams, where detection of the 4.44\,MeV gamma-rays gave a reference time for neutron time-of-flight (TOF) measurements. The PuBe source was a blend of plutonium oxide and beryllium metal sealed in an X.3 capsule which emitted approximately 2.9~$\times$~10$^6$ neutrons per second~\cite{radiochemicalcentre} nearly isotropically. See Ref.~\cite{SCHERZINGER201798} for further details.

\subsection{Detectors}
\label{subsection:ApparatusDetectors}

\begin{figure}[H]
    \centering
    \includegraphics[width=\textwidth]{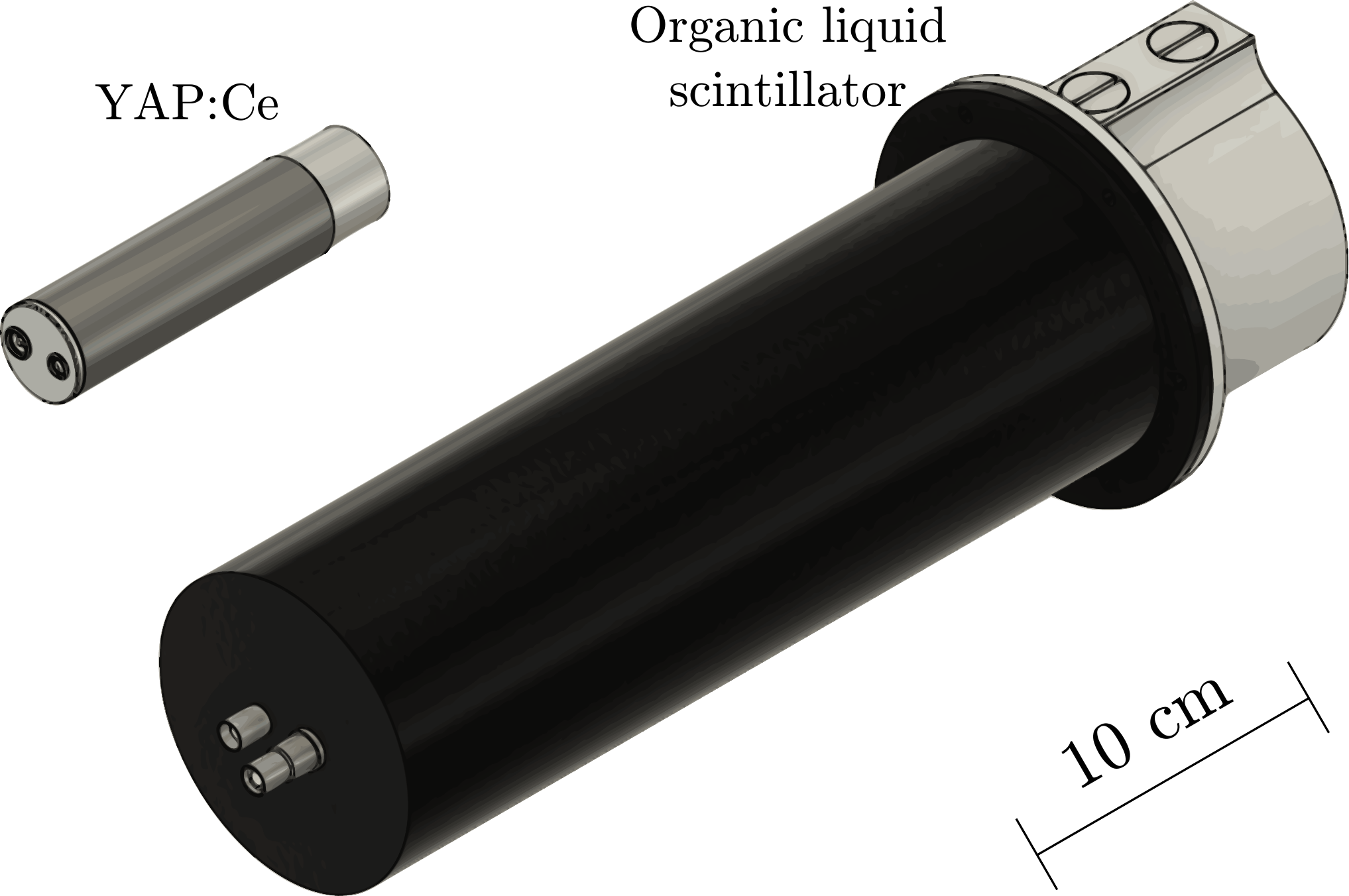}
    \caption{CAD representations of the detectors (to scale). The PMTs lie to the left while the scintillator enclosures lie to the right.
    }
    \label{figure:CAD}
\end{figure}

Figure \ref{figure:CAD} shows CAD representations of two types of scintillation counters employed in this investigation. Cells of organic liquid detected fast neutrons and gamma-rays. They provided the start signal for TOF measurements as well as the pulse-shape information used to discriminate neutrons from gamma-rays (PSD). YAP:Ce crystals detected gamma-rays and provided the stop signal for TOF measurements. 

\subsubsection{Gamma-ray trigger detectors}
\label{subsubsection:ApparatusDetectorsGammaRayTriggerDetectors}

Yttrium Aluminum Perovskit:Cerium (Ce$^{+}$ doped YAlO$_{3}$, YAP:Ce) inorganic crystals~\cite{MOSZYNSKI1998157} have good gamma-ray detection efficiency and low efficiency for neutrons in the energy range investigated here. Four YAP:Ce detectors were used to detect both actinide-cascade and 4.44\,MeV gamma-rays in the presence of the intense fast-neutron field of the PuBe source. Provided by Scionix~\cite{scionix}, they consisted of a cylindrical 1\,in. $\times$  1\,in. (diameter $\times$ height) crystal mounted on a 1\,in. Hamamatsu Type R1924 PMT~\cite{hamamatsu}. A $^{22}$Na gamma-ray source ($E_{\gamma}$~=~1.28\,MeV) was used to set the YAP:Ce gains with HV values of $\sim-$750\,V. The YAP:Ce detectors, with minimal sensitivity to the fast neutrons from the PuBe source, provided the gamma-ray tags.

\subsubsection{Fast-neutron/gamma-ray detector}
\label{subsubsection:ApparatusDetectorsFastNeutronGammaRayDetector}

Cylindrical aluminum cups, 94\,mm in diameter by 62\,mm deep with a thickness of 3\,mm, contained the liquid scintillators. The inside of each cup was painted with EJ~520~\cite{ej520}, a TiO$_2$-based reflector. A 5\,mm thick borosilicate glass window~\cite{borosilicate} was glued to each cup using Araldite 2000$+$~\cite{araldite}. For each detector, the cup/window assembly constituted a $\sim$430\,cm$^{3}$ liquid-scintillator cell. The liquid scintillators were first flushed with nitrogen and then pushed into the cells with a nitrogen gas-transfer system. The entrance/exhaust ports were sealed with screws and Viton O-rings~\cite{viton}. Filled cells were fitted without optical coupling grease (dry fitted) to 57\,mm long by 72.5\,mm diameter cylindrical lightguide made from PMMA UVT~\cite{pmma}. The lightguides were coated externally with EJ~510~\cite{ej510}, a TiO$_2$-based reflector. These cell/lightguide assemblies were dry fitted to 3\,in. diameter Electron Tubes type 9821KB PMTs~\cite{et_9821kb}. The assemblies were placed within mu-metal magnetic shields and a spring was used to hold the cell, PMMA, and PMT faces tightly together. The operating voltage of the detector was set at $-2$\,kV, as employed in previous investigations of similar detectors~\cite{SCHERZINGER201574, JEBALI2015102, JuliusScherzinger2016, SCHERZINGER201798}. 
HV was not tuned to match the gains of the detectors, but variable attenuators (CAEN type N858~\cite{caen_n858}) were inserted  to equalize the signal amplitudes passed to a waveform digitizer. After matching, typical 1\,MeV$_{ee}$ signals had amplitudes of $\sim$700\,mV, with risetimes of $\sim$5\,ns and falltimes of $\sim$60\,ns. The cells were filled with two different liquid scintillators developed for fast-neutron detection. These were NE~213A~\cite{ANNAND1997} (a pseudocumene-based variant of NE~213~\cite{ne213}) possessing excellent fast neutron PSD properties and EJ~305~\cite{ej305}, a pseudocumene-based organic scintillator with one of the highest scintillation-light yield outputs of any liquid scintillator. EJ~305 has a long optical attenuation length but poorer PSD properties compared to NE~213A. Selected scintillator properties are presented in Table~\ref{table:ScintillatorProperties}.

\begin{table}[H]
    \footnotesize
    \centering
    \caption{Selected scintillator properties.}
    \begin{tabular}{r r r}\hline
                         Scintillator &      NE~213A &       EJ~305 \\ \hline
                                 Base & Pseudocumene & Pseudocumene \\
            Flash point ($^{\circ}$C) &     $\sim$54 &     $\sim$45 \\
                   Density (g/cm$^3$) &    $\sim$0.9 &   $\sim$0.89 \\
Pristine light output (\% anthracene) &   $\sim$75\% &   $\sim$80\% \\
              Decay times (short, ns) &      $\sim$3 &    $\sim$2.7 \\
        Peak emission wavelength (nm) &    $\sim$420 &    $\sim$425 \\ 
                            \hline
    \end{tabular}
    \label{table:ScintillatorProperties}
    \end{table} 

\subsection{Configuration}
\label{subsection:ApparatusConfiguration}

\begin{figure}[H]
    \centering
    \includegraphics[width=\textwidth]{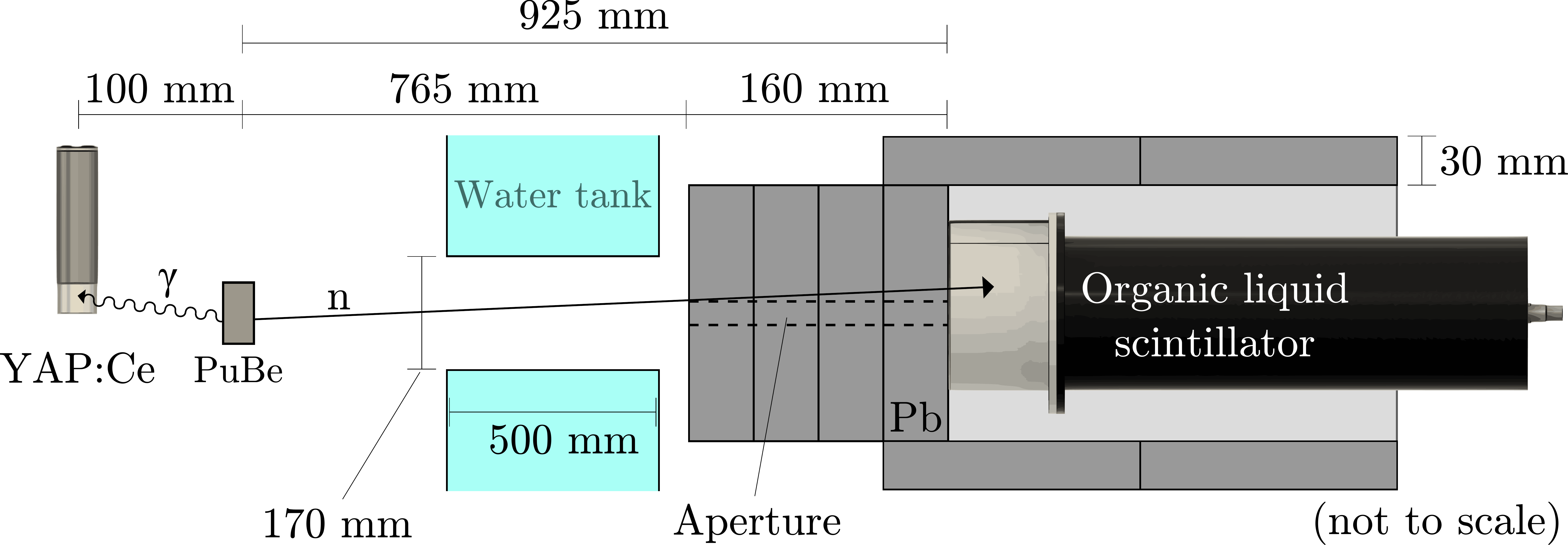}
    \caption{Experimental setup (not to scale). Conceptually, the PuBe source positioned at the center of a water tank emitted correlated 4.44\,MeV gamma-ray/fast-neutron pairs. The neutrons were detected in a Pb-shielded liquid-scintillator detector located outside the water tank to trigger the TOF measurement. The corresponding gamma-rays were detected in a YAP:Ce detector located inside the water tank to stop the TOF measurement. A small aperture was left in the shielding house to facilitate the gamma-flash measurements necessary to calibrate TOF.}
    \label{figure:ExperimentalSetup}
\end{figure}

Figure~\ref{figure:ExperimentalSetup} shows a sketch of the experimental setup. The PuBe source was placed so that its cylindrical-symmetry axis corresponded to the vertical direction in the lab. Four YAP:Ce detectors were also placed with the cylindrical-symmetry axes in the vertical direction, each approximately 10\,cm from the PuBe source, with slightly varying out-of-plane positions. These detectors measured low-energy cascade gamma-rays from $^{234}$U as well as the energetic 4.44\,MeV gamma-rays coming from $\alpha$~$+$~$^{9}$Be~$\rightarrow$~$^{12}$C$^{*}$~$\rightarrow$~n~$+$~$\gamma$. The PuBe source and four YAP:Ce detectors were encased in a water-filled shielding cube known as the ``Aquarium"~\cite{international2020iaea} with a water-filled wall thickness of $\sim$50\,cm. A $\sim$17\,cm diameter penetration in each of the four walls of the Aquarium allowed gamma-rays and fast neutrons to escape the shielding confinement, defining beams. At one beam-port exit, a Pb castle was constructed to encase the liquid-scintillator detectors and facilitate the reproducibility of the positioning of the detector. Room background measured inside the castle using a 1.5\,in. CeBr$_{3}$ detector with a $-$50\,mV threshold was $<$1\,Hz. The NE~213A detector with a $-$25\,mV threshold showed a background rate of $\sim$50\,Hz. As shown in Fig.~\ref{figure:ExperimentalSetup}, the liquid-scintillator detectors were placed in the castle with the upstream face of the cell at a distance of 92.5\,cm from the center of the PuBe source and at source height. The cylindrical symmetry axis of this detector pointed directly at the center of the source. A $\sim$10x10\,mm$^2$ aperture in the 16\,cm thick face of the Pb castle allowed this detector to measure source-related low-energy cascade gamma-rays and energetic 4.44\,MeV gamma-rays. Two types of coincidence events were of special interest: 
\begin{enumerate}
\item{a fast neutron detected in the liquid-scintillator detector in coincidence with a 4.44\,MeV gamma-ray detected in a YAP:Ce detector (a ``tagged-neutron" event).}
\item{prompt, time-correlated gamma-ray pairs emitted from the PuBe source detected in the liquid-scintillator detector and a YAP:Ce detector (a ``gamma-flash" event).}
\end{enumerate}
See Ref.~\cite{SCHERZINGER201574} for further details.

\subsection{Electronics and data acquisition}
\label{subsection:MeasurmentElectronicsAndDataAcquisition}

\begin{figure}[H]
    \centering
    \includegraphics[width=\textwidth]{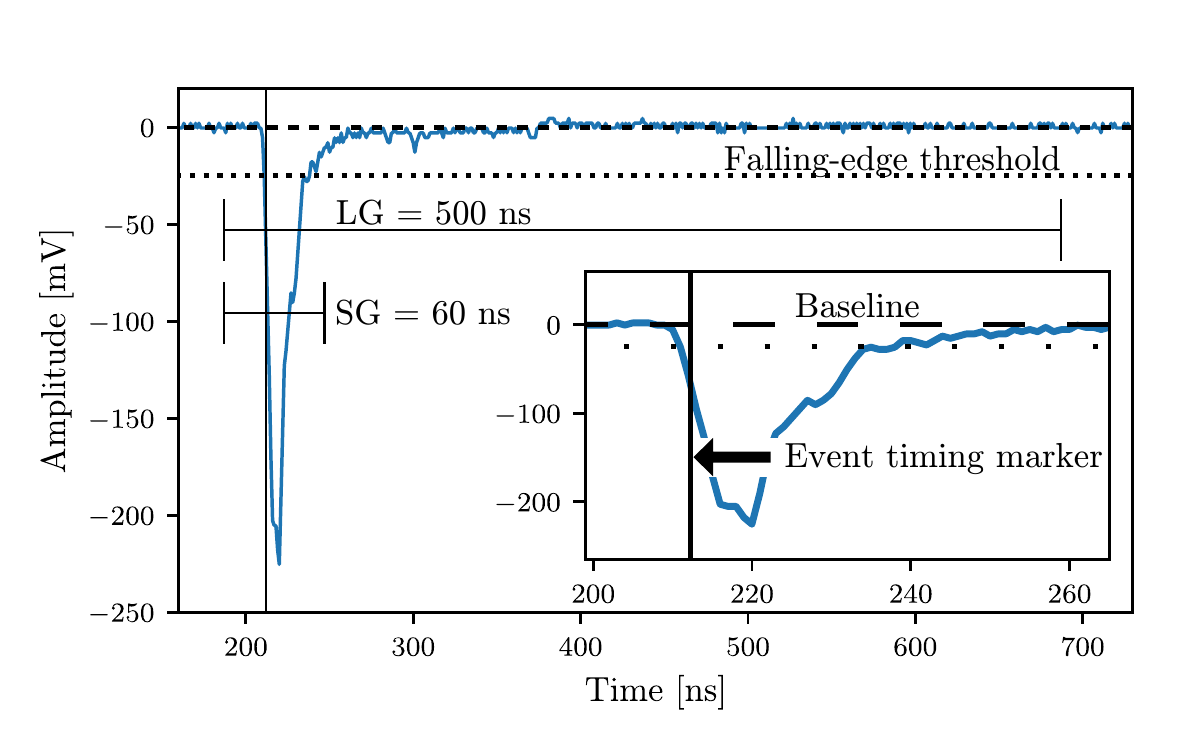}
    \caption{Signal waveform. Typical signals had a $\sim$5\,ns risetime, a $\sim-$230\,mV amplitude, and a $\sim$50\,ns falltime. The falling-edge threshold was set to $-$25\,mV. Also illustrated are the event-timing marker and both the 500\,ns long-gate (LG) and 60\,ns short-gate (SG) integration windows, each opening 25\,ns before this marker.}
    \label{figure:WaveForm}
\end{figure}

A CAEN VX1751 Waveform Digitizer~\cite{caen_vx1751} (10 bit, 500\,MHz analog bandwidth) was employed for data acquisition. The dynamic range was $-$1\,V and 10$^{3}$ samples were taken during the 1\,\textmu{}s acquisition window. A $-$25\,mV internal falling-edge trigger threshold was used.
The software~\cite{nppp2020} employed to analyze the waveform of each pulse included  \python-based~\cite{python} code libraries \pandas~\cite{pandas}, \SciPy~\cite{scipy}, and \texttt{numpy}~\cite{numpy}. The liquid-scintillator event timing was determined using an interpolating zero-crossover method~\cite{GFKNOLL} which reduced the time walk associated with the internal hardware trigger of the digitizer. For each scintillation pulse, the signal baseline was subtracted and the resulting waveform (see Fig.~\ref{figure:WaveForm}) was integrated, effectively giving the signal charge. This charge integration (6.35$\pm$5.5\% fC/channel) was started 25\,ns before the event-timing marker, and the integration gate length was set to 60\,ns for a short-gate (SG) integration and 500\,ns for a long-gate (LG) integration. Pulse shape (PS) was parametrized using the ``tail-to-total" method  ~\cite{JHINGAN2008165,LAVAGNO2010492,PAWELCZAK201321} (Eq.~\ref{equation:PS}) where the difference in the signals registered by the LG and SG integrations was normalized to the signal registered by the LG integration.

\begin{equation}
    \rm
    PS = \frac{LG - SG}{ LG}
    \label{equation:PS}
\end{equation}

\subsection{Energy calibration}
\label{subsection:MeasurementEnergyCalibration}

\begin{figure}[H]
    \centering
    \includegraphics[width=\textwidth]{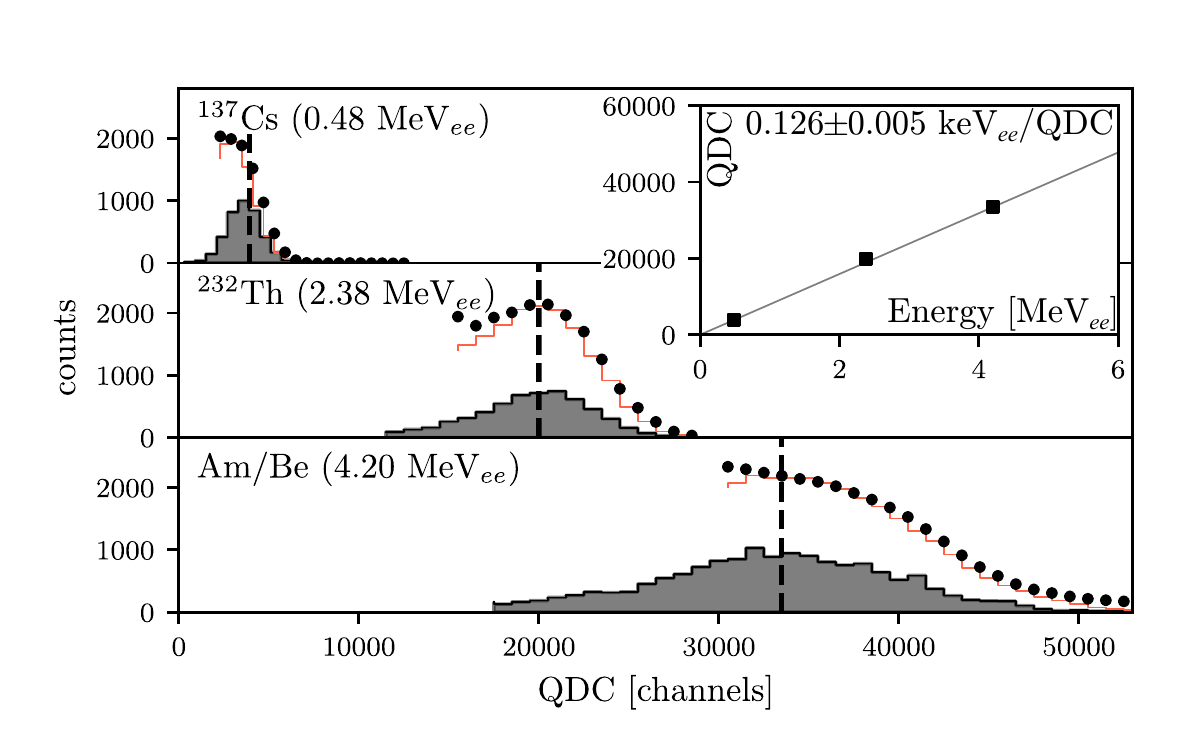}
    \caption{Energy calibration. Measured and simulated Compton distributions for three gamma-ray energies incident on the NE~213A scintillator. Measurement: filled circles; simulation: orange histograms; simulation with tight cut on maximum electron energy: gray shaded histograms. The dashed lines show the mean values of the gray distributions. The insert plots these mean values against the gamma-ray energies along with a linear fit. For interpretation of the references to color in this figure caption, the reader is referred to the web version of this article.}
    \label{figure:EnergyCalibration}
\end{figure}

The scintillation-light yield produced by incident gamma-rays in organic liquids is almost linear above $\sim$100~keV~\cite{KNOX1972519,GFKNOLL}. Below pair-production threshold, Compton scattering dominates because of the low average $Z$ value of the scintillator. Compton edges were measured with the gamma-ray sources listed in Table~\ref{table:GammaRaySources} and the results are displayed in Fig.~\ref{figure:EnergyCalibration}.

\begin{table}[H]
    \centering
    \caption{Gamma-ray sources. The gamma-ray energies and recoiling electron energies at the Compton edges $E_{\rm CE}$ are shown.}
    \begin{tabular}{r c c} \hline
                Source     & $E_{\gamma}$ [MeV] & $E_{\rm CE}$ [MeV$_{ee}$] \\
                $^{137}$Cs &               0.66 &                      0.48 \\
                $^{232}$Th &               2.62 &                      2.38 \\
$^{241}$Am/$^{9}$Be (AmBe)&               4.44 &                      4.20 \\ \hline
    \end{tabular}
    \label{table:GammaRaySources}
\end{table}

Each detector was aligned so that the cylindrical symmetry axis pointed at the calibration source. $^{137}$Cs and $^{232}$Th were placed at a distance of 50\,cm from the face of the detector, while the stronger AmBe was placed at a distance of 200\,cm. A typical run lasted one hour. Count rates were $<$1\,kHz and thus pileup and dead time were negligible. Room background was subtracted from each spectrum using a real-time normalization. Gain drift was corrected for offline. The response of the detectors to the gamma-rays was simulated using \geant~\cite{AGOSTINELLI2003250} version 4.10.04~\cite{1610988} patch~03 (8 February 2019) with a physics list based on the electromagnetic physics classes \texttt{G4EmStandardPhysics}, \texttt{G4EmExtraPhysics} and including the hadronic interaction class \texttt{FTFP\_BERT\_HP}. Gamma-ray response was simulated by tracking secondary electrons and the resulting scintillation photons, which were followed to the photocathode. Here, the wavelength-dependent quantum efficiency~\cite{et_9821kb} gave the probability of a photon producing a photoelectron. The scale factor necessary to match the number of simulated photoelectrons at the photocathode to the measured signal charge was the only free parameter. The positions of the Compton edges were determined from the simulations by placing a tight cut around the electron energy very close to the edge, as decribed in Ref.~\cite{mauritzson2021geant4based}. The result of the energy calibration is presented in Fig.~\ref{figure:EnergyCalibration}. 

\begin{figure}[H]
    \centering
    \includegraphics[width=\textwidth]{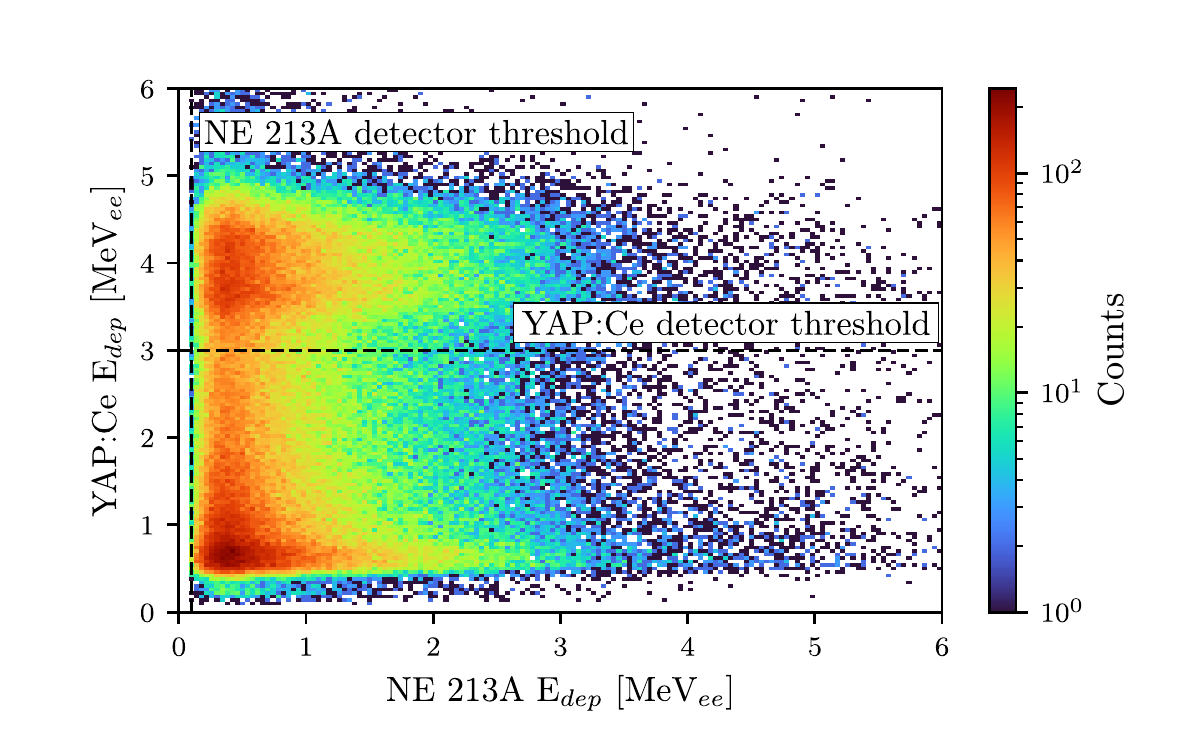}
    \caption{Energy correlation. Energy deposited in the YAP:Ce gamma-ray detector versus energy deposited in NE 213A liquid-scintillator detector. The dashed lines indicate the threshold used for each detector, 100\,keV$_{ee}$ for the NE~213A and 3\,MeV$_{ee}$ for the YAP:Ce. The structure lying above the YAP:Ce threshold corresponds to candidate tagged-neutron events.}
    \label{figure:EnergyCorrelation}
\end{figure}

Figure~\ref{figure:EnergyCorrelation} shows the typical correlation between the energy deposited in the YAP:Ce detector and the energy deposited in the NE~213A liquid-scintillator detector. The YAP:Ce energy calibration employed the Compton edge of the 4.44\,MeV gamma-ray, which does not produce a full-energy peak in the 1 in. crystal, and the full-energy peak from the 1.28\,MeV gamma-ray of $^{22}$Na. The linear fit is constrained to pass through the origin. Events above the YAP:Ce detector threshold of 3\,MeV$_{ee}$ result from the 4.44\,MeV gamma-ray and thus correspond to candidate tagged-neutrons events. This threshold was held fixed during the analysis. Events below the YAP:Ce detector threshold correspond overwhelmingly to lower energy gamma-rays, but may include the occasional neutron (YAP:Ce is largely insensitive to neutrons). The subset of these events which produce a coincident signal in the liquid scintillator are gamma-flash events, corresponding to time-correlated gamma-rays being detected in both detectors.

\section{Results}
\label{section:Results}

\begin{figure}[H]
    \centering
    \includegraphics[width=\textwidth]{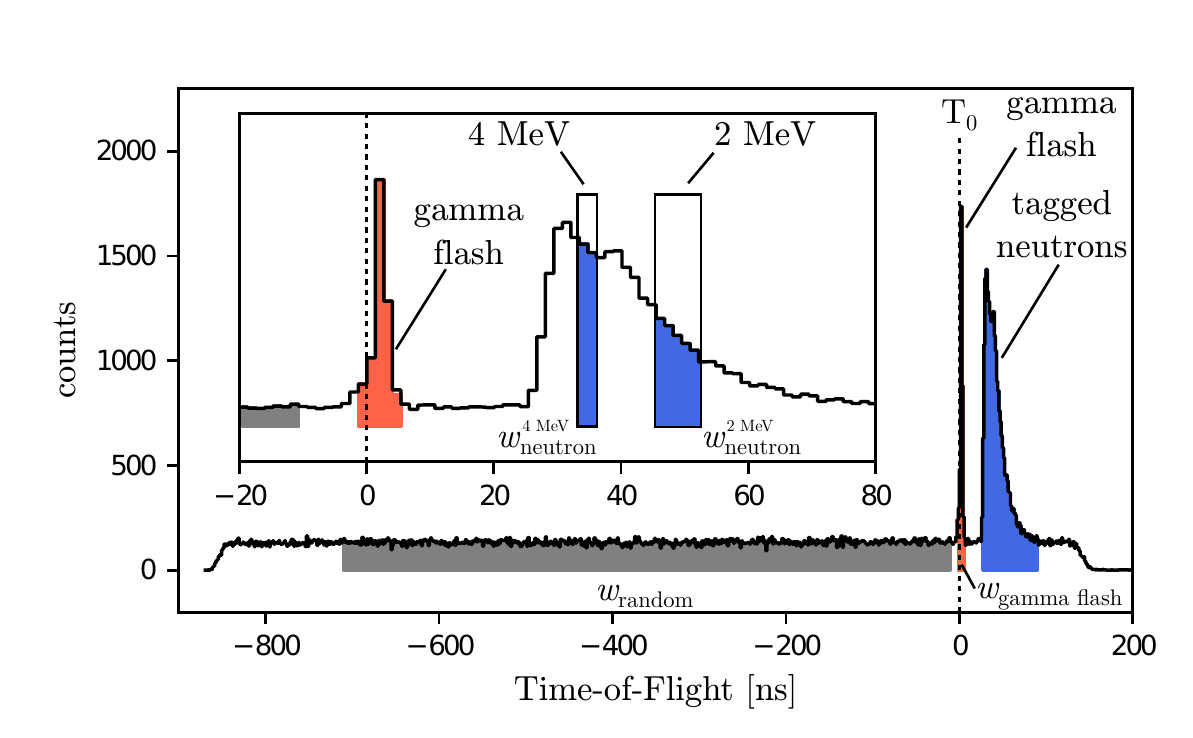}
    \caption{TOF spectrum. The full range of the acquisition window (including the flat random region) and an expanded view of the region of main interest are shown. The boxes illustrate the TOF ranges of 500\,keV wide bins of neutron kinetic energy centered at 2 and 4\,MeV.}
    \label{figure:BackgroundSubtraction}
\end{figure}

Figure~\ref{figure:BackgroundSubtraction} presents a representative TOF distribution obtained for an average neutron drift distance of 96\,cm. The vertical dashed line located at 0\,ns and labeled $T_{0}$ locates the time of particle production in the PuBe source. It is inferred from the gamma flash timing, which is given by the sharp peak to the right of $T_{0}$ at $\sim$2.9\,ns. The gamma flash corresponds overwhelmingly to one of the correlated gamma-rays being detected in a YAP:Ce detector and the other in the liquid-scintillator detector. The $\sim$1\,ns width of the gamma flash is due to a combination of electronic jitter and finite detector volumes. The broader peak to the right of the gamma flash is due to tagged neutrons, and corresponds to a 4.44\,MeV gamma-ray being detected in a YAP:Ce detector while the corresponding fast neutron is detected in the liquid-scintillator detector. The gamma-flash and tagged fast-neutron distributions are clearly separated. A flat distribution of random-coincidence events spans the entire TOF distribution. These random events arise from uncorrelated signals in the YAP:Ce and liquid scintillators which fall randomly within the acquisition window. Since the counting rates were relatively low, the random distribution is flat and and straightforward to subtract from under the correlated distributions~\cite{OWENS1990574}.

\begin{figure}[H]
    \centering
    \includegraphics[width=\textwidth]{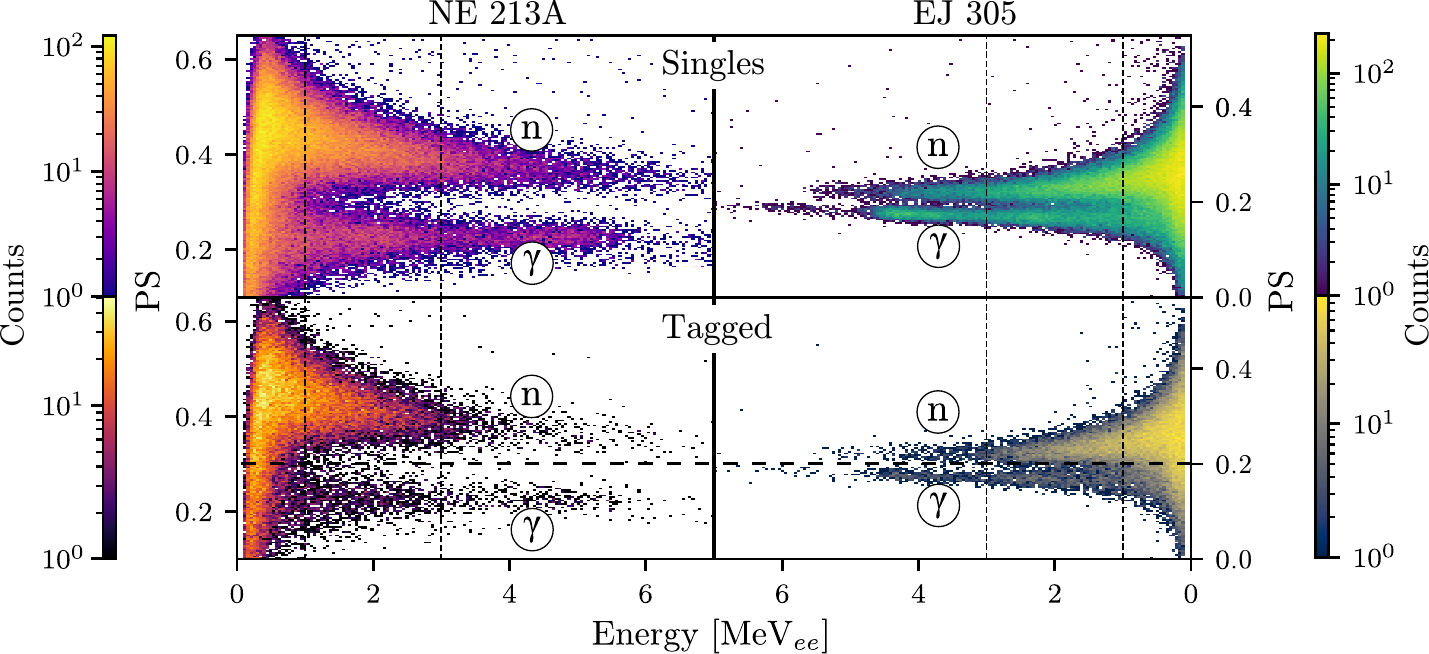}
    \caption{Pulse shape versus energy. A software threshold of 100\,keV$_{ee}$ has been applied to the data from both scintillators. The dashed vertical lines at 1 and 3\,MeV$_{ee}$ are representative of the systematic series of threshold cuts applied in the subsequent data analysis. The long-dashed horizontal line indicates the boundary between neutron (above) and gamma-ray (below) events. Note the y-axes and color scales are different for NE~213A and EJ~305.}
    \label{figure:PS_vs_QDC}
\end{figure}

Figure~\ref{figure:PS_vs_QDC} shows PS versus energy deposited in the liquid-scintillator detectors for singles and tagged data. The tagged data are kinematically restricted to neutron kinetic energies below $\sim$6\,MeV due to energy taken by the 4.44\,MeV gamma-ray. For both scintillators, neutrons result in larger PS values as they produce more of the slower scintillation component. The separation between the neutron distributions and the gamma-ray distributions is considerably larger for NE~213A where the optimum cut for neutron/gamma-ray separation sits at PS~$=$~0.3. For EJ~305, the corresponding cut would be at PS~$=$~0.2. NE~213A clearly provides superior PSD.

\begin{figure}[H]
    \centering
    \includegraphics[width=\textwidth]{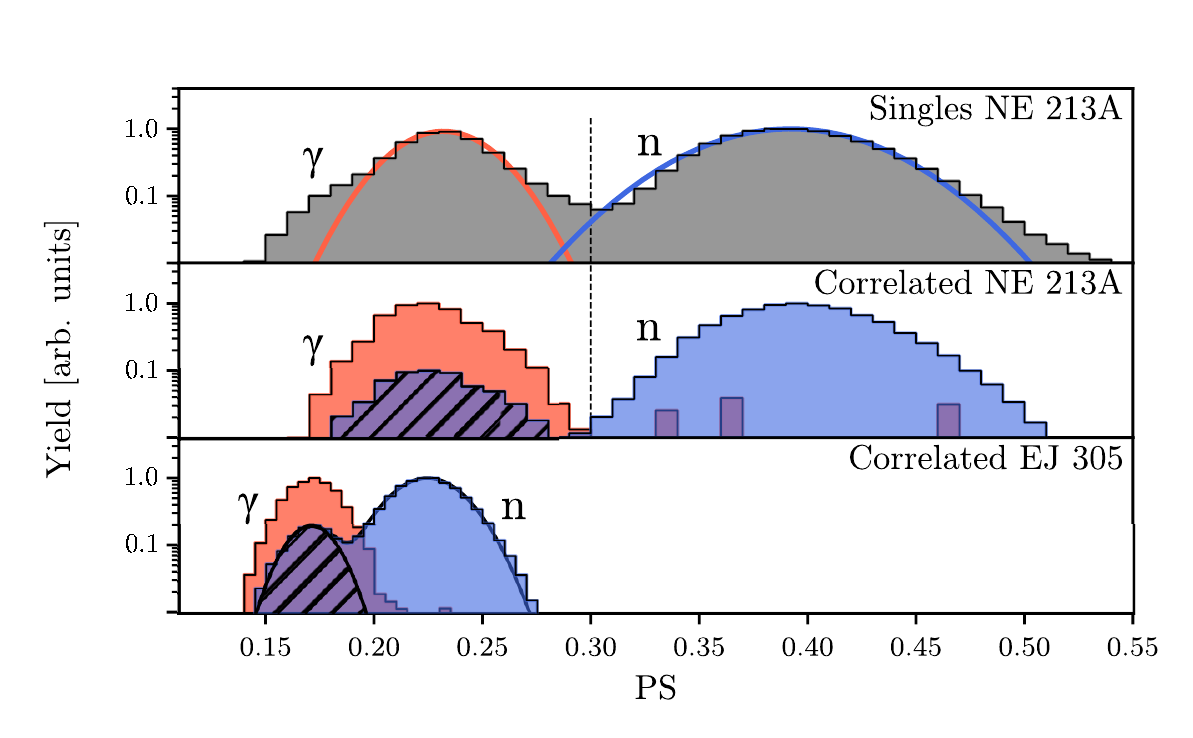}
    \caption{Non-prompt gamma-rays (NPGs). PS spectra for an energy threshold of 2\,MeV$_{ee}$. Top: NE~213A singles data where there is no coincidence with the YAP:Ce detectors. The red and blue curves are Gaussian fits to the gamma-ray and neutron distributions. Middle: correlated NE~213A data where there is a coincidence. Gamma-ray and neutron distributions are completely separated using TOF. Neutron and NPG distributions are separated by PS. Bottom: correlated EJ~305 data where there is a coincidence. The shaded area shows the Gaussian fit used to estimate the NPG contribution. For interpretation of the references to color in this figure caption, the reader is referred to the web version of this article.}
    \label{figure:NPG_gamma_rays}
\end{figure}

Figure~\ref{figure:NPG_gamma_rays} shows PS data from the detectors produced with a 2\,MeV$_{ee}$ detector threshold. Overlap at the 10\% level between the gamma-ray and neutron distributions is displayed for the singles data for NE~213A shown in the top panel. In contrast, much cleaner separation is shown in the tagged, random-subtracted distributions for NE~213A presented in the middle panel. In fact, the particle identification in these correlated results, produced using the tagging technique, is sufficient to separate gamma-rays, neutrons, and non-prompt gamma-rays (NPGs). NPGs are source-related, tagged events which have the non-relativistic TOF signature of a fast neutron but the PS signature of a gamma-ray. They are due to tagged fast neutrons scattering inelastically, predominantly from the lead shielding, or much less frequently from the materials from which the neutron detector was constructed. The resulting de-excitation gamma-rays are then detected in the liquid-scintillator detector. As NPGs are tagged and not random, they appear very clearly in the correlated PS spectra, and a PS$~>~$0.3 cut is sufficient to remove them. As shown in the bottom panel, the separation between gamma-rays and neutrons is substantially smaller for EJ~305, and significant overlap between the distributions exists. Thus a PS cut does not remove NPGs cleanly and a Gaussian fit to the left side of the NPG/neutron distribution has been employed to estimate their contribution.

\noindent
A Canberra~\cite{canberra} model GC2018 HPGe gamma-ray detector with a 7935SL-2 cryostat and 2002CSL preamp was placed inside the Pb shielding house (no aperture) to measure the NPG spectrum. The detected gamma-rays are overwhelmingly due to the de-excitation of low-lying excited states of the isotopes of $^{\rm nat}$Pb, dominated by the 2615\,keV line from the first excited state of $^{208}$Pb. The indicated PS$~>~$0.3 cut was applied to the correlated neutron data obtained with NE~213A to remove the NPG-related contamination.

\begin{figure}[H]
    \centering
    \includegraphics[width=\textwidth]{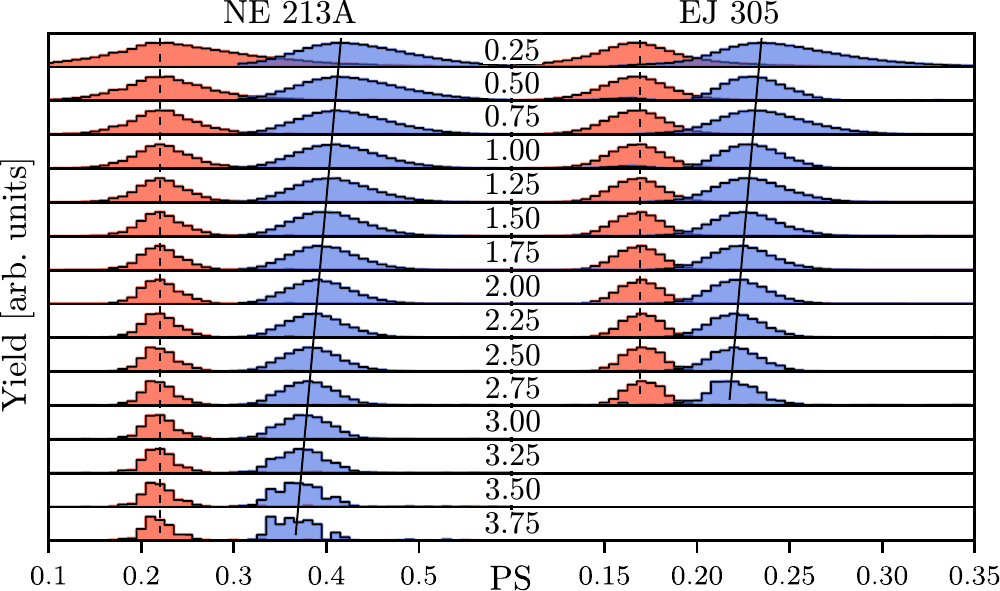}
    \caption{Intrinsic threshold-dependent pulse shape. Correlated gamma-ray distributions are shown in red and neutron distributions are shown in blue. TOF cuts together with PS cuts (NE~213A) or fitting (EJ~305) have been employed to remove NPGs. The threshold in MeV$_{ee}$ for each pad is shown in the center of the plot. Lines show linear fits to the means of the distributions, dashed for gamma-rays and solid for neutrons. For interpretation of the references to color in this figure caption, the reader is referred to the web version of this article.}
    \label{figure:PSvsThreshold}
\end{figure}

Figure~\ref{figure:PSvsThreshold} shows the PS distributions for the correlated data from NE~213A and EJ~305, for a software threshold that has been increased from 0.25 to 3.75\,MeV$_{ee}$ in steps of 0.25\,MeV$_{ee}$. The gamma-ray distributions have been scaled to match the heights of the neutron distributions for purposes of visualization. NPGs have been identified using TOF and PS. For NE~213A, they have been removed with a PS$~>~$0.3 cut. For EJ~305 the NPG contribution was estimated from a Gaussian fit (Fig.~ \ref{figure:NPG_gamma_rays}) and subtracted. The tagging approach facilitates the unambiguous identification of gamma-rays and neutrons. The gamma-ray peak locations do not change as the threshold is varied and sit at PS~$\sim$0.22 for NE~213A and $\sim$0.18 for EJ~305. The peak locations of the neutron PS distributions decrease linearly with increasing threshold, so that the separation between the gamma-ray and neutron peak locations is largest for the lowest threshold and decreases as the applied threshold increases. However the widths of the PS distributions also decrease with increasing threshold, so that neutron/gamma-ray separation actually improves significantly.

\begin{figure}[H]
    \centering
    \includegraphics[width=\textwidth]{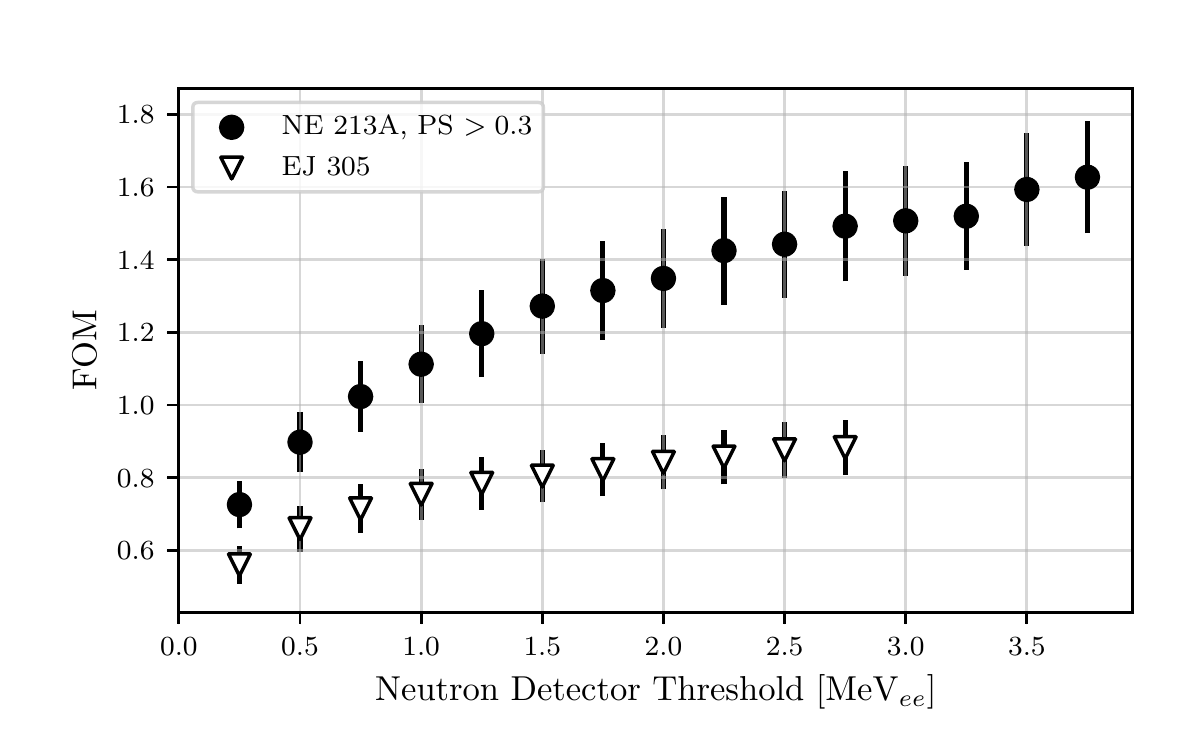}
    \caption{Threshold-dependent figures-of-merit. Numerically evaluated intrinsic FOM resulting from the correlated data presented in Fig.~\ref{figure:PSvsThreshold} for NE~213A (black dots) and EJ~305 (open triangles). Derivation of the uncertainties is detailed in the text.}
    \label{figure:FOMvsThreshold}
\end{figure}

The Figure-of-Merit (FOM) used to characterize the quality of PSD in organic scintillators is commonly defined as
\begin{equation}
    \text{FOM} = \frac{\mid\mu_n-\mu_\gamma\mid}{\delta_n+\delta_\gamma},
    \label{eq:FOM}
\end{equation}
\noindent
where $\mu_{n,\gamma}$ and $\delta_{n,\gamma}$ are the mean positions and Full-Widths-at-Half-Maximum (FWHM) of the gamma-ray and neutron PS distributions respectively. The larger the FOM, the better the PSD. Figure~\ref{figure:FOMvsThreshold} shows the intrinsic FOM as a function of software threshold. The excellent separation of the gamma-ray and neutron PS distributions facilitated a numerical analysis of peak locations and widths over $\pm$3$\sigma$ of the respective distributions. The uncertainties in the resulting FOM values were established from a error analysis which considered both linear and quadratic propagation of the uncertainties from the numerical approach, with the relative uncertainty in the standard deviation was estimated to be $\sim$5\%. For both NE~213A and EJ~305, the uncertainties were $\sim$10\%. For both NE~213A and EJ~305, the intrinsic FOM values increase with increasing detector threshold. NE~213A demonstrates superior FOM for all thresholds. 

\begin{figure}[H]
    \centering
    \includegraphics[width=\textwidth]{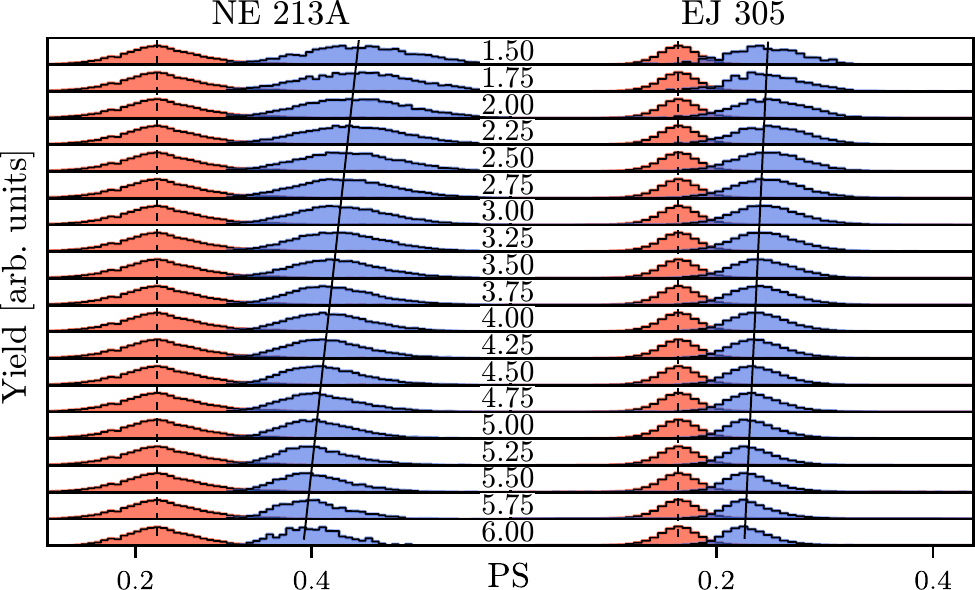}
    \caption{Intrinsic neutron kinetic-energy dependent pulse shape. Correlated gamma-ray distributions are shown in red and neutron distributions are shown in blue. TOF cuts together with PS cuts (NE~213A) or fitting (EJ~305) have been employed to remove NPGs. The neutron kinetic energies in MeV ($\pm0.125$\,MeV) for each plot are shown in the center of the figure. The lines show linear fits to the mean values of the distributions, dashed for gamma-rays and solid for neutrons. For interpretation of the references to color in this figure caption, the reader is referred to the web version of this article.} 
    \label{figure:PSvsTOF}
\end{figure}

Figure~\ref{figure:PSvsTOF} shows the evolution of PS spectra for the correlated gamma-ray and neutron data for both NE~213A and EJ~305 as the average neutron kinetic energy obtained from TOF is raised from 1.50 to 6.00\,MeV in 0.25\,MeV steps. A 0.5\,MeV$_{ee}$ software threshold was employed. The gamma-ray distributions have been scaled to match the heights of the neutron distributions for purposes of visualization. NPGs have been identified using TOF and PS and removed or corrected for as described previously. The peak locations of the gamma-ray distributions do not change with neutron kinetic energy and sit at PS~$\sim$0.22 for NE~213A and $\sim$0.18 for EJ~305. Again, the peak locations of the neutron distributions decrease linearly with increasing neutron kinetic energy. The separation between the gamma-ray and neutron peaks is largest for the lowest energy and decreases as the kinetic energy increases, but this is offset by the increase in distribution widths as energy decreases.

\begin{figure}[H]
    \centering
    \includegraphics[width=\textwidth]{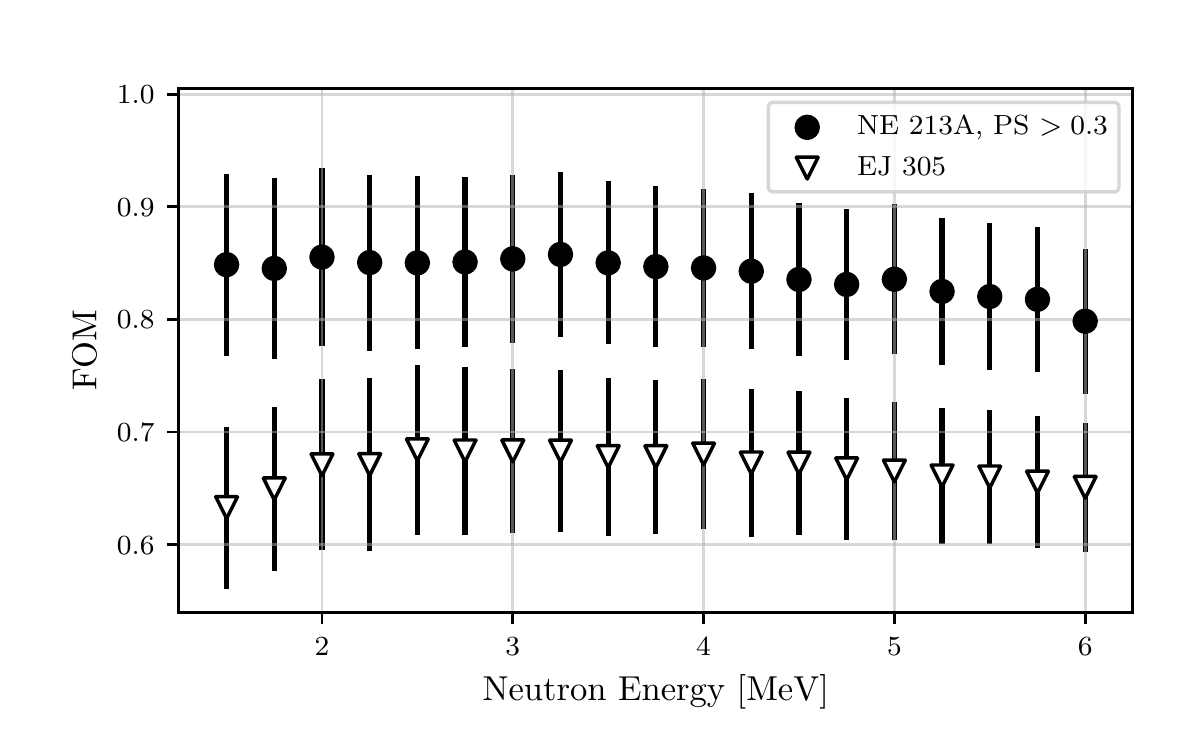}
    \caption{Energy-dependent figures-of-merit. FOM extracted from the correlated data presented in Fig.~\ref{figure:PSvsTOF} for NE~213A (black dots) and EJ~305 (open triangles). Derivation of the uncertainties is detailed in the text.}
    \label{figure:FOMvsTn}
\end{figure}

Figure~\ref{figure:FOMvsTn} shows the intrinsic FOM values extracted from the data displayed in Fig.~\ref{figure:PSvsTOF} as a function of neutron kinetic energy using the same methods as for the data of Figs.~\ref{figure:PSvsThreshold}~and~\ref{figure:FOMvsThreshold}. For both NE~213A and EJ~305, the uncertainties were $\sim$10\%. The neutron-energy dependencies of the FOM distributions are quite similar, but clearly the NE~213A values are considerably larger than EJ~305.

\section{Summary and Discussion}
\label{Section:SummaryAndDiscussion}

A systematic quantification of neutron/gamma-ray PSD has been performed for the organic liquid scintillators NE~213A and EJ~305 using energy-tagged neutrons from $\sim$1.5~--~6\,MeV provided by a PuBe source. Tagging relies on the $\alpha$~$+$~$^{9}$Be~$\rightarrow$~$^{12}$C$^*$~$\rightarrow$~$n$~$+$~$\gamma$ reaction, where the neutron and gamma-ray are detected in coincidence and the gamma-rays provide a time reference for neutron TOF measurement. YAP:Ce crystals were used to detect these 4.44\,MeV gamma-rays, and liquid-scintillator cells were employed to detect the corresponding neutrons (Fig.~\ref{figure:CAD}). The source and YAP:Ce detectors were located inside a water-filled shielding cube, which had cylindrical penetrations to allow for the passage of fast neutrons. Pb was used to shield the liquid scintillators from the direct gamma-ray field from the source and also the 2.22\,MeV gamma-rays resulting from neutron capture by H in the water (Fig.~\ref{figure:ExperimentalSetup}). Event-by-event waveform digitization of the detector signals (Fig.~\ref{figure:WaveForm}) was employed. Measured spectra were calibrated using the Compton edges from gamma-ray sources, whose pulse-height positions were established using a Geant4-based simulation (Fig.~\ref{figure:EnergyCalibration}). The correlation between the gamma-ray energy in the YAP:Ce detector and the neutron energy deposited in the liquid-scintillator cells (Fig.~\ref{figure:EnergyCorrelation}) facilitated the selection of neutron-tagging events. Time-of-flight was employed to determine neutron energy. The measurement time window of length 1\,$\mu$s allowed accurate determination of random-coincidence background, which was then subtracted (Fig.~\ref{figure:BackgroundSubtraction}). PS was evaluated using the tail-to-total method and resulted in excellent separation of gamma-ray and neutron distributions (Fig.~\ref{figure:PS_vs_QDC}) when neutron tagging with random subtraction and removal of NPGs (Fig. ~\ref{figure:NPG_gamma_rays}) was employed. Untagged PS distributions show significantly poorer neutron/gamma-ray separation.

The evolution of the correlated gamma-ray and neutron PS distributions was investigated as the software threshold was raised from 0.25 to 3.75\,MeV$_{ee}$ (Fig.~\ref{figure:PSvsThreshold}). The resulting FOM values (Fig.~\ref{figure:FOMvsThreshold}) show clearly that NE~213A has superior PSD to EJ~305. NE~213A FOM values improve significantly as the pulse-height threshold is increased, while EJ~305 showed a weaker dependence of FOM on threshold. This level of detail for the PSD performance is only possible when the extra information provided by the tagging technique is obtained. The correlated gamma-ray and neutron PS distributions were also investigated for neutron kinetic energy (TOF) cuts (Fig.~\ref{figure:PSvsTOF}) ranging from 1.50 to 6.00\,MeV. While both NE~213A and EJ~305 demonstrated FOM values that are more or less constant as a function of energy (Fig.~\ref{figure:FOMvsTn}), NE~213A is again clearly superior to EJ~305.  

With the PuBe source, the tagged-energy range is relatively small and extension would require an accelerator-based neutron generator. While the neutron-tagging technique offers increased insight into threshold-dependent scintillator response, a clear advantage of the technique lies in the measurement of scintillator response as a function of incident neutron energy using polychromatic neutron sources. Such information is crucial, for example, to the validation of simulation efforts.

\section*{Acknowledgements}
\label{Section:Acknowledgements}
Support for this project was provided by the European Union via the Horizon 2020 BrightnESS Project (Proposal ID 676548) and the UK Science and Technology Facilities Council (Grant No. ST/P004458/1).
\newpage

\bibliographystyle{elsarticle-num}

\begin{thebibliography}{00}

\bibitem{ne213} 
NE~213 is no longer produced. Eljen Technologies EJ-301 (\url{https://eljentechnology.com/products/liquid-scintillators/ej-301-ej-309} [Accessed 2022, Jun. 1] or Saint Gobain BC-501 (\url{https://www.crystals.saint-gobain.com/products/bc-501a-bc-519} [Accessed 2022, Jun. 1] are very similar.

\bibitem{ANNAND1997} 
J.R.M.~Annand et al.,
Nucl. Instr. and Meth. in Phys. Res. A. 400, (1997) 344.
\href{https://doi.org/10.1016/S0168-9002(97)01021-8}{doi:10.1016/S0168-9002(97)01021-8.}

\bibitem{BATCHELOR196170} 
R.~Batchelor et al.,
Nucl. Instr. and Meth. 13, (1961) 70.
\href{https://doi.org/10.1016/0029-554X(61)90171-9}{doi:10.1016/0029-554X(61)90171-9.}

\bibitem{BAYAT2012217} 
E.~Bayat et al.,
Rad. Phys. and Chem. 81, (3), (2012) 217.
\href{https://doi.org/10.1016/j.radphyschem.2011.10.016}{doi:10.1016/j.radphyschem.2011.10.016.}

\bibitem{IWANOWSKA201334} 
J.~Iwanowska et al.,
Nucl. Instr. and Meth. in Phys. Res. A. 712, (2013) 34.
\href{https://doi.org/10.1016/j.nima.2013.01.064}{doi:10.1016/j.nima.2013.01.064.}

\bibitem{PAWELCZAK201321} 
I.A. Pawełczak et al.,
Nucl. Instr. and Meth. in Phys. Res. A. 711, (2013) 21.
\href{https://doi.org/10.1016/j.nima.2013.01.028}{doi:10.1016/j.nima.2013.01.028.}

\bibitem{JEBALI2015102} 
R.~Jebali et al.,
Nucl. Instr. and Meth. in Phys. Res. A. 794, (2015) 102.
\href{https://doi.org/10.1016/j.nima.2015.04.058}{doi:10.1016/j.nima.2015.04.058.}

\bibitem{ej305} 
EJ~305 Highest Light output Liquid Scintillator, 
\url{http://www.ggg-tech.co.jp/maker/eljen/ej-305.html} [Accessed 2022, Jun. 1].

\bibitem{nudat} 
NuData 3.0, National Nuclear Data Center, Brookhaven National Laboratory, 
\url{http://www.nndc.bnl.gov/nudat2/} [Accessed 2022, Jun. 1].

\bibitem{radiochemicalcentre} 
Exactly 4.26~$\times$~10$^6$ neutrons per second. Calibration certified at The Radiochemical Centre, Amersham, England HP7 9LL on 3 September 1973.

\bibitem{SCHERZINGER201798} 
J.~Scherzinger et al.,
Appl. Radiat. Isop. 127 (2017) 98.
\href{https://doi.org/10.1016/j.apradiso.2017.05.014}{doi:10.1016/j.apradiso.2017.05.014.}

\bibitem{MOSZYNSKI1998157} 
M.~Moszyński et al,
Nucl. Instr. and Meth. 101 (1972) 519.
\href{https://doi.org/10.1016/S0168-9002(97)01115-7}{doi:10.1016/S0168-9002(97)01115-7.}

\bibitem{scionix} 
Scionix Holland BV,
\url{http://www.scionix.nl} [Accessed 2022, Jun. 1].

\bibitem{hamamatsu} 
Hamamatsu Photonics, 
\url{http://www.hamamatsu.com} [Accessed 2022, Jun. 1].

\bibitem{ej520} 
EJ~520 reflective paint, 
\url{http://www.ggg-tech.co.jp/maker/eljen/ej-520.html} [Accessed 2022, Jun. 1].

\bibitem{borosilicate} 
Borofloat, \url{https://www.schott.com/en-us/products/borofloat} [Accessed 2022, Jun. 1], supplied by Glasteknik i Emmaboda AB, Utv\"{a}gen 6 SE-361 31 Emmaboda, Sweden.

\bibitem{araldite} 
Araldite is a registered trademark of Huntsman, \url{http://www.araldite2000plus.com} [accessed 2022, Jun. 1].

\bibitem{viton} 
Viton is a registered trademark of DuPont Performance Elastomers LLC.

\bibitem{pmma} 
Poly-methyl-methacrylate, also known as PMM, acrylic, plexiglass, and lucite. Supplied by Nordic Plastics Group AB, Bronsyxegatan 6, SE-213 75 Malmoe, Sweden.

\bibitem{ej510} 
EJ~510 reflective paint,
\url{https://eljentechnology.com/products/accessories/ej-510} [Accessed 2022, Jun. 1].

\bibitem{et_9821kb} 
Electron Tubes type 9821KB, \url{https://et-enterprises.com/products/photomultipliers/product/p9821b-series} [Accessed 2022, Jun. 1].

\bibitem{SCHERZINGER201574} 
J.~Scherzinger et al.,
Appl. Radiat. Isop. 98 (2015) 74.
\href{https://doi.org/10.1016/j.apradiso.2015.01.003}{doi:10.1016/j.apradiso.2015.01.003.}

\bibitem{JuliusScherzinger2016} 
J.~Scherzinger et al.,
Nucl. Instr. and Meth. in Phys. Res. A. 840, (2016) 121.
\href{https://doi.org/10.1016/j.nima.2016.10.011}{doi:10.1016/j.nima.2016.10.011.}

\bibitem{SCHERZINGER2017128} 
J.~Scherzinger et al.,
Appl. Radiat. Isop. 128 (2017) 270.
\href{https://doi.org/10.1016/j.apradiso.2017.05.022}{doi:10.1016/j.apradiso.2017.05.022.}

\bibitem{caen_n858} 
CAEN N858 dual attenuator, \url{https://www.caen.it/products/n858/} [Accessed 2022, Jun. 01].

\bibitem{international2020iaea} 
Modern Neutron Detection,
TECDOC Series No. 1935, International Atomic Energy Agency, Vienna, Austria (2020)
ISBN: 978-92-0-126520-3.

\bibitem{caen_vx1751} 
CAEN VX1751 Waveform Digitizer, \url{https://www.caen.it/products/vx1751/} [Accessed 2022, Jun. 01].

\bibitem{nppp2020} 
H.~Perrey et al., Nuclear Physics Pulse Processing Library,
available at \url{https://gitlab.com/ANPLU/nppp} [Accessed 2022, Jun. 01].

\bibitem{python} 
G. van Rossum and F.L. Drake (editors), Python Reference Manual, PythonLabs (2001).
Python 3.8.5 available at \url{https://www.python.org/} [Accessed 2022, Jun. 01].

\bibitem{pandas} 
The Pandas Development Team,
\texttt{Pandas} 1.2.3 available at \url{https://pandas.pydata.org/}, \href{https://doi.org/10.5281/zenodo.4572994}{doi:10.5281/zenodo.4572994.} [Accessed 2022, Jun. 01]

\bibitem{scipy} 
\texttt{SciPy} 1.6.1 available at \url{https://www.scipy.org/}, \href{https://doi.org/10.5281/zenodo.4547611}{doi:10.5281/zenodo.4547611.} [Accessed 2022, Jun. 01]

\bibitem{numpy} 
\texttt{numpy} 1.20.3 available at \url{https://pypi.org/project/numpy/} [Accessed 2022, Jun. 01].

\bibitem{GFKNOLL} 
Radiation detection and measurement,
G.F.~Knoll,
2nd edition, Wiley, New York, U.S.A. (1989) 222,
ISBN: 9780471815044.

\bibitem{JHINGAN2008165} 
A.~Jhingan et al.,
Nucl. Instr. and Meth. in Phys. Res. A. 585, (2008) 165.
\href{https://doi.org/10.1016/j.nima.2007.11.013}{doi:10.1016/j.nima.2007.11.013.}

\bibitem{LAVAGNO2010492} 
A.~Lavagno et al.,
Nucl. Instr. and Meth. in Phys. Res. A. 617, (2010) 492.
\href{https://doi.org/10.1016/j.nima.2009.10.111}{doi:10.1016/j.nima.2009.10.111.}

\bibitem{KNOX1972519} 
H.H.~Knox et al.,
Nucl. Instr. and Meth. 101 (1972) 519.
\href{https://doi.org/10.1016/0029-554X(72)90040-7}{doi:10.1016/0029-554X(72)90040-7.}

\bibitem{AGOSTINELLI2003250} 
S.~Agostinelli et al.,
Nucl. Instr. and Meth. in Phys. Res. A. 506, (2003) 250.
\href{https://doi.org/10.1016/S0168-9002(03)01368-8}{doi:10.1016/S0168-9002(03)01368-8.}

\bibitem{1610988} 
J.~Allison et al.,
IEEE Trans. Nucl. Sci. 53, (2006) 270.
\href{https://doi.org/10.1109/TNS.2006.869826}{doi:10.1109/TNS.2006.869826.}

\bibitem{mauritzson2021geant4based} 
N.~Mauritzson et al.,
Nucl. Instr. and Meth. in Phys. Res. A. 1023, (2022) 165962.
\href{https://doi.org/10.1016/j.nima.2021.165962}{doi:10.1016/j.nima.2021.165962.}

\bibitem{OWENS1990574} 
R.O.~Owens,
Nucl. Instr. and Meth. in Phys. Res. A. 288, (1990) 574.
\href{https://doi.org/10.1016/0168-9002(90)90154-X}{doi:10.1016/0168-9002(90)90154-X.}

\bibitem{canberra} 
Canberra model GC2018 HPGe, \url{https://www.canberra.com} [Accessed 2022, Jun. 01].

\end{thebibliography}

\end{document}